\documentclass[twocolumn,twocolappendix,times]{aastex6}
\usepackage{jab,amsmath}
\hypersetup{citecolor=black,linkcolor=black}
\slugcomment{}
\shorttitle{Magnetosheath Turbulence}
\shortauthors{Chen \& Boldyrev}

\begin{document}
\title{Nature of Kinetic Scale Turbulence in the Earth's Magnetosheath}
\author{C.~H.~K.~Chen\altaffilmark{1}, S.~Boldyrev\altaffilmark{2,3}}
\affil{$^1$Department of Physics, Imperial College London, London SW7 2AZ, UK; christopher.chen@imperial.ac.uk\\
$^2$Department of Physics, University of Wisconsin--Madison, Madison, WI 53706, USA\\
$^3$Space Science Institute, Boulder, CO 80301, USA}
\begin{abstract}
We present a combined observational and theoretical analysis to investigate the nature of plasma turbulence at kinetic scales in the Earth's magnetosheath. In the first decade of the kinetic range, just below the ion gyroscale, the turbulence was found to be similar to that in the upstream solar wind: predominantly anisotropic, low-frequency and kinetic Alfv\'en in nature. A key difference, however, is that the magnetosheath ions are typically much hotter than the electrons, $T_\mathrm{i}\gg T_\mathrm{e}$, which, together with $\beta_\mathrm{i}\sim 1$, leads to a change in behaviour in the second decade, close to electron scales. The turbulence here is characterised by an increased magnetic compressibility, following a mode we term the \emph{inertial kinetic Alfv\'en wave}, and a steeper spectrum of magnetic fluctuations, consistent with the prediction $E_B(k_\perp)\propto k_\perp^{-11/3}$ that we obtain from a set of nonlinear equations. This regime of plasma turbulence may also be relevant for other astrophysical environments with $T_\mathrm{i}\gg T_\mathrm{e}$, such as the solar corona, hot accretion flows, and regions downstream of collisionless shocks.
\end{abstract}

\section{Introduction}

Plasma turbulence is widespread, occurring in a variety of astrophysical environments, such as galaxy clusters, the interstellar medium, and stellar winds. It can be modelled as a cascade of energy from large scales, where the energy is injected, to small, kinetic scales (comparable to the ion and electron gyroscales), where it is thought to be dissipated. However, many aspects of how the cascade operates (in particular, at kinetic scales), remain to be understood.

The majority of observations of turbulence at kinetic scales are from the solar wind upstream of the Earth's bow shock \citep[e.g.,][]{bruno13,alexandrova13a,goldstein15,chen16b}. At scales smaller than the ion gyroscale, the energy spectrum of solar wind density and magnetic fluctuations is observed to be a power law in wavenumber, close to $k^{-2.8}$, down to electron scales \citep[e.g.,][]{alexandrova09,kiyani09a,chen10b,chen12a,sahraoui13a}. Between ion and electron scales, the fluctuations are anisotropic, with stronger gradients perpendicular to the mean magnetic field than parallel, $k_\perp\gg k_\|$ \citep{chen10b}, and the amplitude of the density fluctuations relative to the magnetic fluctuations indicates that the turbulence is predominantly low frequency, with a polarisation consistent with that of the kinetic Alfv\'en wave \citep{chen13c,boldyrev13a}. These features can be interpreted as a cascade of kinetic Alfv\'en turbulence from ion to electron scales \citep{howes08a,schekochihin09,chen10a,boldyrev12b}.

When the critical balance principle \citep{goldreich95} is assumed, in which the linear and nonlinear terms of the dynamical equations are comparable, dimensional arguments lead to a perpendicular energy spectrum $E(k_\perp)\propto k_\perp^{-7/3}$, along with an anisotropy $k_\|\propto k_\perp^{1/3}$ \citep{cho04,schekochihin09}. If the cascade is also assumed to accumulate into intermittent 2D structures, these scalings become $E(k_\perp)\propto k_\perp^{-8/3}$ and $k_\|\propto k_\perp^{2/3}$ \citep{boldyrev12b}. The critical balance principle suggests that the linear physics is relevant, even in a strongly turbulent, intermittent cascade. This is borne out in the solar wind, with a variety of observations showing that the fluctuations follow linear relationships to order unity \citep[e.g.,][]{bale05,sahraoui09,howes10,he11b,podesta11d,yao11,howes12a,klein12,tenbarge12b,chen13a,chen13c,kiyani13,bruno15,telloni15,verscharen17}, and similarly in numerical simulations \citep[e.g.,][]{howes11a,verscharen12a,tenbarge12a,boldyrev12b,franci15a,told15,cerri16}, although there can also be quantitative differences introduced by the nonlinearites \citep[e.g.,][]{boldyrev11,boldyrev12a,chen13b,chen13c}.

The Earth's magnetosheath, the region of solar wind downstream of the bow shock, presents a different environment in which kinetic scale turbulence can be measured, although it has been less comprehensively studied here. This is partly because there are often additional processes taking place which complicate the picture, e.g., instability generated waves (such as mirror modes, ion cyclotron waves, and whistler waves) and various other non-turbulent structures \citep[see, e.g.,][for a review]{lucek05}. However, with careful data selection, these can be avoided, allowing the pure turbulent cascade to be investigated. It has been shown that at ion scales, the magnetic field spectrum in the magnetosheath steepens \citep[e.g.,][]{dudokdewit96,czaykowska01,chaston08b,alexandrova08c,yordanova08,yao11b,huang14,stawarz16,matteini17}, the electric field spectrum flattens \citep{chaston08b,stawarz16,matteini17}, and the turbulence in the kinetic range is intermittent \citep{dudokdewit96,sundkvist07,voros16,stawarz16} and anisotropic, with $k_\perp\gg k_\|$ \citep{mangeney06,alexandrova08c,stawarz16}. These features are similar to the upstream solar wind, which might suggest that a similar type of turbulence is present.

An important difference between the magnetosheath and the upstream solar wind, however, is the ratio of ion and electron temperatures. In the upstream solar wind, the temperatures are typically comparable, $T_\mathrm{i}\sim T_\mathrm{e}$, whereas in the magnetosheath, the ions are typically much hotter, $T_\mathrm{i}\gg T_\mathrm{e}$ \citep{wang12}, as a result of processing by the bow shock \citep[e.g.,][]{burgess13,krasnoselskikh13,vink15}. We have found that this can lead to a change in the behaviour of the turbulence near electron scales. In this paper, we present the identification of this turbulence regime in measurements from the \emph{Magnetospheric Multiscale} (\emph{MMS}) spacecraft, together with a theoretical framework through which it can be understood.

\section{Observations}
\label{sec:observations}

\subsection{Data Interval}

The observational analysis is based on data from the four \emph{MMS} spacecraft \citep{burch16} during a period in the Earth's magnetosheath (16th October 2015, 09:24:11--09:25:24) for which burst mode data is available. During this time, the spacecraft were located 11.9\,$R_\mathrm{E}$ from Earth, close to the dusk side magnetopause, with an inter-spacecraft separation of $\sim$\,14\,km. This period does not contain large amplitude variations in the magnetic field magnitude (associated with mirror modes) or high frequency wave packets at electron scales (thought to be whistler waves), allowing the pure turbulent cascade to be studied.

\begin{figure}
\includegraphics[width=\columnwidth,trim=0 0 0 0,clip]{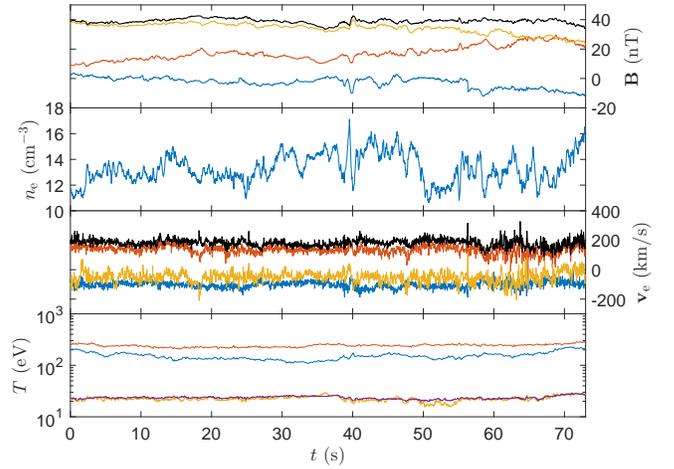}
\caption{Time series of the magnetic field ($\mathbf{B}$) components in GSE (blue, orange, yellow) and magnitude (black), electron number density ($n_\mathrm{e}$), electron velocity ($\mathbf{v}_\mathrm{e}$) components in GSE (blue, orange, yellow) and magnitude (black), and temperatures $T_{\|\mathrm{i}}$ (blue), $T_{\perp\mathrm{i}}$ (orange), $T_{\|\mathrm{e}}$ (yellow) and $T_{\perp\mathrm{e}}$ (purple) from \emph{MMS3}.}
\label{fig:timeseries}
\end{figure}

For the magnetic field $\mathbf{B}$, data from the FGM \citep{russell16} and SCM \citep{lecontel16} instruments were combined using a wavelet technique \citep{chen10b} to produce 8192 samples/s data containing the full range of frequencies (with the crossover between instruments at $\sim$ 8\,Hz). The electric field $\mathbf{E}$ was measured by the SDP \citep{lindqvist16} and ADP \citep{ergun16} instruments at the same resolution. FPI \citep{pollock16} was used for the ion and electron densities $n_\mathrm{i}$ and $n_\mathrm{e}$, velocities $\mathbf{v}_\mathrm{i}$ and $\mathbf{v}_\mathrm{e}$, and temperatures $T_\mathrm{i}$ and $T_\mathrm{e}$; the resolution of these moments is 150\,ms for the ions and 30\,ms for the electrons. A time series of the data from \emph{MMS3} for this interval is shown in Figure \ref{fig:timeseries}. The average plasma conditions were: $B\approx 39$\,nT, $n_\mathrm{i}\approx n_\mathrm{e}\approx14$\,cm$^{-3}$, $v_\mathrm{i}\approx v_\mathrm{e}\approx180$\,km\,s$^{-1}$, $T_\mathrm{i}\approx210$\,eV, $T_\mathrm{e}\approx23$\,eV, with temperature anisotropies $(T_\perp/T_\|)_\mathrm{i}\approx 1.6$ and $(T_\perp/T_\|)_\mathrm{e}\approx 1.0$. These parameters result in average ion and electron plasma betas $\beta_\mathrm{i}\approx0.79$ and $\beta_\mathrm{e}\approx0.087$ (where $\beta_s=2\mu_0n_sk_BT_s/B^2$).

From the magnitude of the ion and electron betas, $\beta_\mathrm{e}\ll\beta_\mathrm{i}\sim 1$, it can be seen that for the ions, the gyroradius $\rho_\mathrm{i}=v_{\mathrm{th}\perp\mathrm{i}}/\Omega_\mathrm{i}$ (where $v_{\mathrm{th}\perp\mathrm{i}}=\sqrt{2k_\mathrm{B}T_{\perp\mathrm{i}}/m_\mathrm{i}}$ is the thermal speed, $\Omega_\mathrm{i}=q_\mathrm{i}B/m_\mathrm{i}$ is the gyrofrequency, $m_\mathrm{i}$ is the mass, and $q_\mathrm{i}$ is the charge) and inertial length $d_\mathrm{i}=v_\mathrm{A}/\Omega_\mathrm{i}$ (where $v_\mathrm{A}=B/\sqrt{\mu_0\rho}$ is the Alfv\'en speed and $\rho$ is the mass density) are similar $\rho_\mathrm{i}\sim d_\mathrm{i}$, whereas for the electrons, the gyroradius is much smaller, $\rho_\mathrm{e}\ll d_\mathrm{e}$. This results in two sub-ranges between the ion and electron gyroscales: one above the electron inertial scale, $1/\rho_\mathrm{i}<k<1/d_\mathrm{e}$, and one below, $1/d_\mathrm{e}<k<1/\rho_\mathrm{e}$. The following sections describe the nature of the fluctuations in each of these ranges. 

\subsection{Nature of Fluctuations at $kd_\mathrm{e}<1$}
\label{sec:kdesmall}

\begin{figure}
\includegraphics[width=\columnwidth,trim=0 0 0 0,clip]{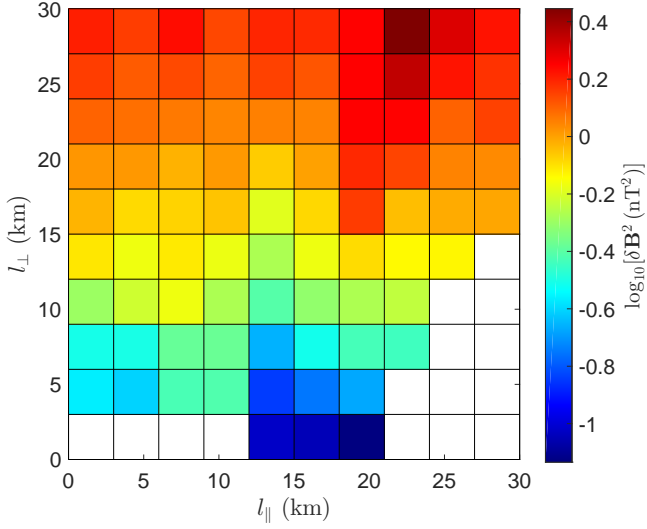}
\caption{Magnetic fluctuation energy $\delta\mathbf{B}^2$ as a function of length scale parallel $l_\|$ and perpendicular $l_\perp$ to the local mean field. Contours are elongated in the $l_\|$ direction, indicating anisotropic fluctuations $k_\perp\gg k_\|$.}
\label{fig:anisotropy}
\end{figure}

To determine the nature of the fluctuations, first the anisotropy was measured, using the multi-spacecraft technique described by \citet{chen10b}. Two-point structure functions $\delta\mathbf{B}^2\left(\mathbf{l}\right)=\left<|\mathbf{B}\left(\mathbf{x}+\mathbf{l}\right)-\mathbf{B}\left(\mathbf{x}\right)|^2\right>_\mathbf{x}$ were calculated from the time-lagged magnetic field measurements between pairs of spacecraft. The technique mixes spatial and temporal measurements, and assumes the Taylor hypothesis \citep{taylor38} to be satisfied (i.e., that the measured temporal variations correspond to spatial variations in the plasma frame), which appears to be the case, despite the low flow speed and dispersive regime (see Section \ref{sec:taylor}). Figure \ref{fig:anisotropy} shows $\delta\mathbf{B}^2\left(\mathbf{l}\right)$, binned and averaged as a function of length scale parallel and perpendicular to the local mean field. The range of scales covered is $11<k\rho_\mathrm{i}<57$, or equivalently $0.29<kd_\mathrm{e}<1.5$. It can be seen that the contours of $\delta\mathbf{B}^2$ are elongated in the parallel direction, and that the value of $\delta\mathbf{B}^2$ at a scale of 15\,km is $\sim$ 10 times larger in the perpendicular direction than in the parallel direction. This indicates strongly anisotropic fluctuations $k_\perp\gg k_\|$, consistent with previous findings for magnetosheath \citep{mangeney06,alexandrova08c} and solar wind \citep{chen10b} turbulence in the kinetic range.

\begin{figure}
\includegraphics[width=\columnwidth,trim=0 0 0 0,clip]{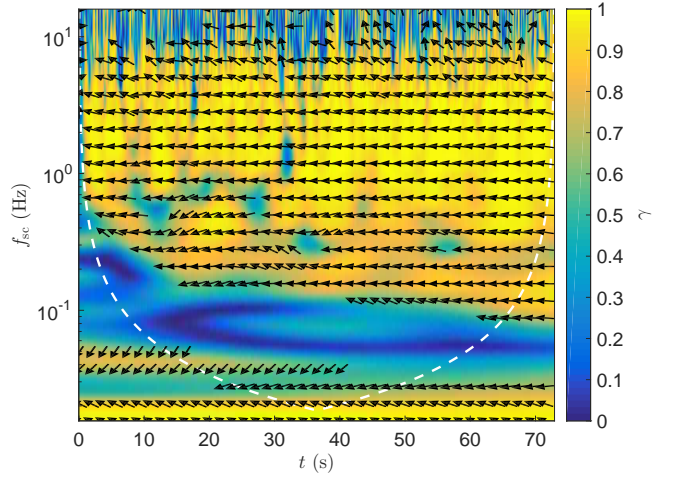}
\caption{Wavelet magnitude-squared coherence $\gamma$ and phase $\phi$ (angle of black arrows from $+x$ direction) between $\delta n_\mathrm{e}$ and $\delta B_\|$. The white dashed line marks the cone of influence. Strong anti-correlation can be seen at spacecraft-frame frequencies $0.5\,\mathrm{Hz}\lesssim f_\mathrm{sc}\lesssim5\,\mathrm{Hz}$.}
\label{fig:correlation}
\end{figure}

The two possible modes in this regime for an isotropic Maxwellian plasma are the kinetic Alfv\'en wave,
\begin{equation}
\omega^2=\frac{k_\|^2v_\mathrm{A}^2k_\perp^2\rho_\mathrm{i}^2}{\beta_\mathrm{i}+2/(1+T_\mathrm{e}/T_\mathrm{i})},
\label{eq:kaw}
\end{equation}
and the oblique whistler wave,
\begin{equation}
\omega^2=k_\|^2v_\mathrm{A}^2k_\perp^2d_\mathrm{i}^2.
\end{equation}
To distinguish these, the correlation between $\delta n$ and $\delta B_\|$ can be used, which is negative for the kinetic Alfv\'en wave and positive for the whistler wave. Figure \ref{fig:correlation} shows the magnitude-squared wavelet coherence, $\gamma$, between $\delta n_\mathrm{e}$ and $\delta B_\|$, and the phase lag $\phi$ (black arrows) for $\gamma>0.5$, measured by \emph{MMS3}. To avoid complications with the definition of $\mathbf{B}_0$, $\delta|\mathbf{B}|$ was used as a proxy for $\delta B_\|$, which requires $\delta\mathbf{B}^2/B_0^2\ll2\delta B_\|/B_0$, a condition well-satisfied here. For spacecraft-frame frequencies $0.5\,\mathrm{Hz}\lesssim f_\mathrm{sc}\lesssim5\,\mathrm{Hz}$, corresponding to $1\lesssim k_\perp\rho_\mathrm{i}\lesssim10$, $\gamma\sim 1$ and there is a strong anti-correlation. The average phase lag in this range is $\left<\phi\right>=(172\pm7)^\circ$ (where the uncertainty is the standard deviation). For $f_\mathrm{sc}\gtrsim 10\,\mathrm{Hz}$, the anti-correlation is lost due to noise in the density measurement. This strong anti-correlation, along with the $k_\perp\gg k_\|$ anisotropy, indicates the predominantly kinetic Alfv\'en nature of the turbulence in the first decade of the kinetic range.

\begin{figure}
\includegraphics[width=\columnwidth,trim=0 0 0 0,clip]{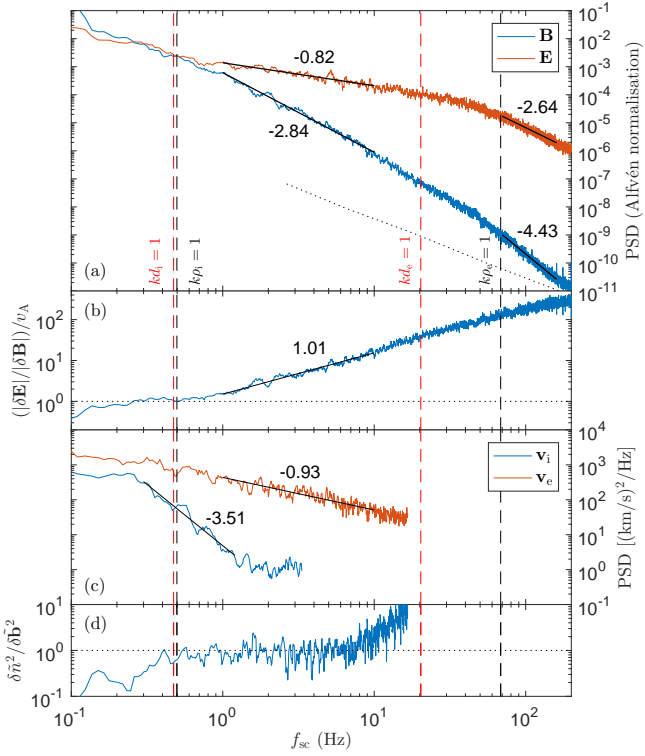}
\caption{(a) Magnetic field, $\mathbf{B}$, and electric field, $\mathbf{E}$, power spectra. (b) Ratio of electric and magnetic fluctuations. (c) Ion velocity, $\mathbf{v}_\mathrm{i}$, and electron velocity, $\mathbf{v}_\mathrm{e}$, power spectra. (d) Ratio of density and magnetic fluctuations (Equation (\ref{eq:kawnorm})). The dashed lines mark the plasma microscales under the Taylor hypothesis; the dotted line is the normalized SCM noise floor \citep{lecontel16}.}
\label{fig:spectra}
\end{figure}

Figure \ref{fig:spectra} shows the spectra of various quantities measured by \emph{MMS3}, calculated using the multitaper method \citep{percival93}. In the range $1/\rho_\mathrm{i}<k<1/d_\mathrm{e}$, the trace magnetic fluctuation spectrum has a power law index of $-2.8$ before steepening at electron scales, similar to solar wind observations \cite[e.g.,][]{alexandrova09,kiyani09a,chen10b,chen12a,sahraoui13a} and kinetic Alfv\'en turbulence simulations \citep[e.g.,][]{howes11a,boldyrev12b}, and not far from the $k_\perp^{-8/3}$ prediction for intermittent kinetic Alfv\'en turbulence \citep{boldyrev12b}. The trace electric field spectrum, Lorentz transformed into the zero mean velocity frame \citep{chen11b}, has a spectral index of $-0.8$, a factor of $k^2$ shallower than the magnetic spectrum, before also steepening at electron scales. The ratio $(|\delta\mathbf{B}|/|\delta\mathbf{E}|)/v_\mathrm{A}$ is around unity for $k\rho_\mathrm{i}<1$ then displays linear scaling for $k\rho_\mathrm{i}>1$, as also seen by \citet{matteini17}. This is because kinetic Alfv\'en fluctuations, although electromagnetic in nature, have a significant potential component of the electric field; the ion motion satisfies $\omega\ll k_\perp v_{\mathrm{th},\mathrm{i}}$, so the density adjusts to the electric potential as $\delta n\propto\phi$, and since $\delta n\propto\delta B$, the electric field is given by $E(k)\approx-ik\phi(k) \propto kB(k)$. The electron velocity spectral index is $-0.9$, close to $k^2$ shallower than the magnetic field and similar to the electric field, since the electron velocity is dominated by the $\mathbf{E}\times\mathbf{B}_0$ drift. The ion velocity spectrum, however, is much steeper, similar to in the upstream solar wind \citep{safrankova13a,safrankova16}, reaching the instrumental noise around $k\rho_\mathrm{i}\approx3$. This is because the ions no longer participate in the same $\mathbf{E}\times\mathbf{B}_0$ drift as the electrons, since their gyroradius is much larger than the scales of the electric field fluctuations in this range. Figure \ref{fig:spectra}(d) shows the normalized ratio of density and magnetic fluctuations,
\begin{equation}
\frac{\delta\tilde{n}^2}{\delta\tilde{\mathbf{b}}^2}=\frac{\beta_\mathrm{i}}{2}\left(1+\frac{T_\mathrm{e}}{T_\mathrm{i}}\right)\left[1+\frac{\beta_\mathrm{i}}{2}\left(1+\frac{T_\mathrm{e}}{T_\mathrm{i}}\right)\right]\frac{\left(\delta n/n_0\right)^2}{\left(\delta\mathbf{B}/B_0\right)^2},
\label{eq:kawnorm}
\end{equation}
which is $\delta\tilde{n}^2/\delta\tilde{\mathbf{b}}^2\sim 1$, confirming that the turbulence is predominantly low frequency, $\omega\ll k_\perp v_{\mathrm{th},\mathrm{i}}$, and kinetic Alfv\'en rather than whistler \citep{chen13c,boldyrev13a}. The increase in $\delta\tilde{n}^2/\delta\tilde{\mathbf{b}}^2$ for $f_\mathrm{sc}\gtrsim10\,\mathrm{Hz}$ is not physical, but due to the density spectrum reaching the noise level.

\subsection{Nature of Fluctuations at $kd_\mathrm{e}\gtrsim1$}
\label{sec:kdelarge}

While the first decade of the magnetosheath kinetic range, $1\lesssim k\rho_\mathrm{i}\lesssim10$, described in Section \ref{sec:kdesmall}, is similar to that in the upstream solar wind, the nature of the turbulence changes in the second decade. The kinetic Alfv\'en wave (Equation (\ref{eq:kaw})) is derived assuming $\omega\ll k_\| v_{\mathrm{th},\mathrm{e}}$, however, it can be seen that (for $T_\mathrm{i}\gg T_\mathrm{e}$) this breaks down at the scale
\begin{equation}
k_\perp^2\rho_\mathrm{e}^2\sim \beta_\mathrm{i}\left(2+\beta_\mathrm{i}\right)\left(T_\mathrm{e}/T_\mathrm{i}\right)^2,
\label{eq:ikawscale}
\end{equation}
about halfway between $k_\perp\rho_\mathrm{i}\sim1$ and $k_\perp\rho_\mathrm{e}\sim1$ for the measured parameters. In the range $k_\| v_{\mathrm{th},\mathrm{e}}\ll\omega\ll k_\perp v_{\mathrm{th},\mathrm{i}}$, the kinetic Alfv\'en wave transforms into a mode with dispersion relation
\begin{equation}
\omega^2=\frac{k_\|^2v_\mathrm{A}^2k_\perp^2\rho_\mathrm{i}^2}{\beta_\mathrm{i}(1+k_\perp^2d_\mathrm{e}^2)(1+2/\beta_\mathrm{i}+k_\perp^2d_\mathrm{e}^2)},
\label{eq:ikaw}
\end{equation}
which we call the \emph{inertial kinetic Alfv\'en wave} (see Appendix \ref{sec:ikawderivation} for the derivation). This should be distinguished from the standard inertial Alfv\'en wave derived under the conditions $\beta_\mathrm{e}\ll m_\mathrm{e}/m_\mathrm{i}$ and $\beta_\mathrm{i}\ll 1$ \cite[e.g.,][]{lysak96}, which are not satisfied here. Note also that the term ``inertial kinetic Alfv\'en wave" has occasionally been applied to the standard inertial Alfv\'en wave \citep{shukla09,agarwal11}, although the conditions assumed in these works are essentially the same as in \citet{lysak96} and different from those leading to Equation (\ref{eq:ikaw}).

The key observational feature of the transition to inertial kinetic Alfv\'en turbulence is the magnetic compressibility,
\begin{equation}
\frac{\delta B_\|^2}{\delta B_\perp^2}=\frac{1+k_\perp^2d_\mathrm{e}^2}{1+2/\beta_\mathrm{i}+k_\perp^2d_\mathrm{e}^2}.
\label{eq:ikawcompressibility}
\end{equation}
For $k_\perp^2d_\mathrm{e}^2\ll 1$, $\delta B_\|^2/\delta B_\perp^2=1/(1+2/\beta_\mathrm{i})$, and for $k_\perp^2d_\mathrm{e}^2\gg1+2/\beta_\mathrm{i}$, $\delta B_\|^2/\delta B_\perp^2=1$, i.e., in general, the magnetic compressibility increases as energy cascades through $k_\perp d_\mathrm{e}\sim1$. This is because for $k_\perp d_\mathrm{e}<1$, the ion pressure reduces plasma compressibility, which due to pressure balance causes a $\beta_\mathrm{i}$-dependent reduction in $\delta B_\|$. For $k_\perp d_\mathrm{e}>1$, however, the compressibility caused by the electron polarization drift becomes stronger than the compressibility due to the E$\times$B drift, leading to $\delta B_\|^2/\delta B_\perp^2=1$.

\begin{figure}
\includegraphics[width=\columnwidth,trim=0 0 0 0,clip]{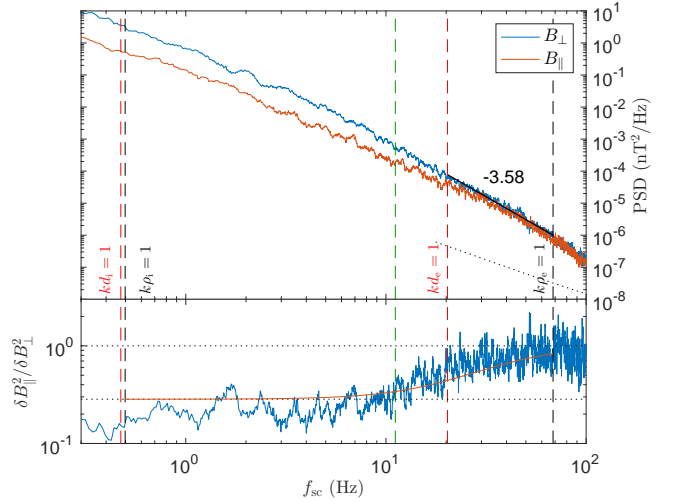}
\caption{Power spectra of $B_\perp$ and $B_\|$ and magnetic compressibility $\delta B_\|^2/\delta B_\perp^2$. In the lower panel, the black dotted lines show the asymptotic predictions $1/(1+2/\beta_\mathrm{i})$ and 1, and the red solid line is Equation (\ref{eq:ikawcompressibility}). Vertical dashed lines are the same as in Figure \ref{fig:spectra}; the additional green line is the transition scale (Equation (\ref{eq:ikawscale})).}
\label{fig:compressibility}
\end{figure}

Figure \ref{fig:compressibility} shows the spectra of $B_\perp$ and $B_\|$, and the ratio of these, as measured by \emph{MMS3}. As for Section \ref{sec:kdesmall}, $\delta|\mathbf{B}|$ was used as a proxy for $\delta B_\|$, and $\delta B_\perp^2=\delta\mathbf{B}^2-\delta B_\|^2$. It can be seen that at $k\rho_\mathrm{i}\sim 1$, the magnetic compressibility is $\approx1/(1+2/\beta_\mathrm{i})$, then this increases through $kd_\mathrm{e}\sim 1$, becoming $\approx1$ by the time $k\rho_\mathrm{e}\sim 1$ is reached. In fact, the red line shows the measured compressibility to be following Equation (\ref{eq:ikawcompressibility}) over the whole range between the ion and electron gyroscales. Note that the increase in compressibility is not due to noise; random fluctuations would produce equal power in all three components, resulting in $\delta B_\|^2/\delta B_\perp^2=0.5$ (which occurs only for $f_\mathrm{sc}>200$\,Hz, where the noise level is reached). It is also not due to parallel propagating whistler wave packets, which would appear as enhancements in the spectrum with a strong circular polarisation \citep{matteini17}, but are not present in this interval. The observed magnetic compressibility, therefore, is consistent with a transition to inertial kinetic Alfv\'en turbulence for $kd_\mathrm{e}\gtrsim 1$.

\subsection{Validity of the Taylor Hypothesis}
\label{sec:taylor}

The use of the Taylor hypothesis to interpret Figures \ref{fig:anisotropy}-\ref{fig:compressibility} in the spatial domain requires careful consideration. This assumes that the plasma-frame frequencies are small compared to the frequency of the convected spatial variations $|\omega|\ll|\mathbf{k}\cdot\mathbf{v}|$. If the fluctuations follow the kinetic Alfv\'en wave (Equation (\ref{eq:kaw})) and inertial kinetic Alfv\'en wave (Equation (\ref{eq:ikaw})) dispersion relations, which the measurements of Sections \ref{sec:kdesmall} and \ref{sec:kdelarge} are consistent with, this condition (for $T_\mathrm{i}\gg T_\mathrm{e}$) becomes $k_\|\rho_\mathrm{i}\ll(v/v_\mathrm{A})\sqrt{\beta_\mathrm{i}(1+k_\perp^2d_\mathrm{e}^2)(1+2/\beta_\mathrm{i}+k_\perp^2d_\mathrm{e}^2)}$ \citep[see also][]{howes14b}. For $k_\perp^2d_\mathrm{e}^2\ll1$, this reduces to $k_\|\rho_\mathrm{i}\ll(v/v_\mathrm{A})\sqrt{2+\beta_\mathrm{i}}\approx 1.3$. Due to the anisotropy $k_\perp\gg k_\|$, this can be well-satisfied, even for $k_\perp\rho_\mathrm{i}\gg1$. For $k_\perp^2d_\mathrm{e}^2\gg 1+2/\beta_\mathrm{i}$ this reduces to $k_\|\rho_\mathrm{i}\ll(v/v_\mathrm{A})\sqrt{\beta_\mathrm{i}}k_\perp^2d_\mathrm{e}^2$, which can be written $k_\|/k_\perp\ll k_\perp v/\Omega_\mathrm{e}$. Therefore, as long as $k_\|/k_\perp$ does not grow faster than $k_\perp$ (theoretical considerations suggest it actually grows slower; see Section \ref{sec:spectrumanisotropy}), the Taylor condition would remain valid down to $k_\perp\rho_\mathrm{e}\sim 1$. 

\begin{figure}
\includegraphics[width=\columnwidth,trim=0 0 0 0,clip]{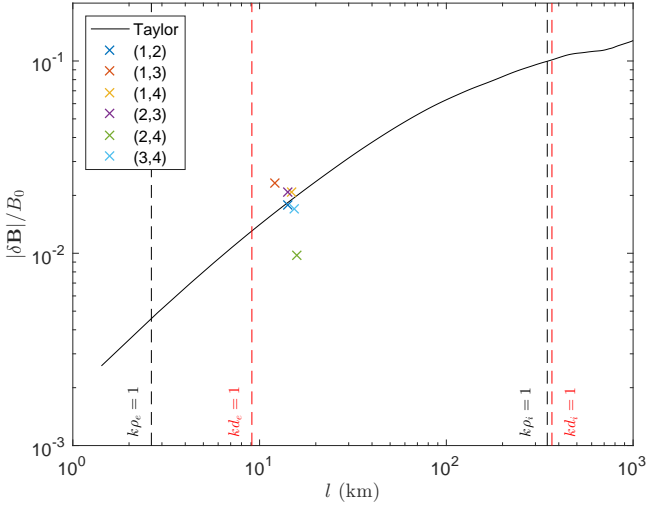}
\caption{Normalized magnetic fluctuation amplitude $|\delta\mathbf{B}|/B_0$ as a function of scale $l$ measured with the single-spacecraft method assuming the Taylor hypothesis (black line), and with pairs of spacecraft $(i,j)$ as a direct spatial measurement. Vertical dashed lines are the same as in Figure \ref{fig:spectra}.}
\label{fig:taylor}
\end{figure}

A recent numerical study \citep{klein14b} concluded that the Taylor hypothesis is indeed satisfied for kinetic Alfv\'en turbulence at $v/v_\mathrm{A}\sim1$, and if it was violated, significantly shallower spectra would result. The fact that the spectra in Figure \ref{fig:spectra} match those in solar wind, as well as expectations from simulations and theory, is consistent with the interpretation that the measured fluctuations are spatial. As a more direct test, Figure \ref{fig:taylor} shows the normalized magnetic fluctuation amplitudes, as calculated from the second order structure function, from both the single-spacecraft method (converting from the temporal to spatial domain assuming the Taylor hypothesis) and as direct spatial measurements from the six pairs of \emph{MMS} spacecraft. There is some scatter in the multi-spacecraft measurements, due to the spacecraft separation vectors being at different angles to the mean field direction, but the average value of $|\delta\mathbf{B}|/B_0=0.018$ is similar to the single-spacecraft value of $|\delta\mathbf{B}|/B_0=0.019$, consistent with the Taylor hypothesis being valid down to $kd_\mathrm{e}\sim 1$.

\section{Theoretical Model}

\subsection{Dynamical Equations for Inertial Kinetic Alfv\'en Turbulence}

To understand the nonlinear properties of the cascade, we first derive the dynamical equations for inertial kinetic Alfv\'en turbulence. We consider strongly anisotropic ($k_\perp\gg k_\|$) fluctuations, and use the ordering $k_\|/k_\perp\sim\delta B/B_0\sim\delta n/n_0\sim\omega/\Omega_\mathrm{e}\ll 1$.

The starting equations are the electron continuity and momentum equations. In the electron continuity equation,
\begin{equation}
\frac{\partial n}{\partial t}+\nabla_\perp \cdot (n{\bf v}_\perp)+\nabla_\| (n v_\|)=0,
\label{eq:econteq}
\end{equation}
the parallel electron velocity can be expressed through the parallel electron current, $v_\|=-J_\|/(ne)$, which can, in turn, be expressed through the $z$ component of the magnetic vector potential, $\psi\equiv-A_z\approx-A_\|$, as $J_\|=(1/\mu_0)\nabla_\perp^2 \psi$. Here, $e$ is the magnitude of the electron charge. Due to the small amplitude of the magnetic fluctuations, $\delta B/B_0\ll 1$, the deviation of the magnetic field lines from the $z$ direction (large-scale mean field) is small, so the parallel components of the vector fields can be approximated by their $z$ components. This is, however, not true for the parallel gradients, since $k_\|\ll k_\perp$, which are given by
\begin{equation}
\nabla_\|=\partial/\partial z+B_0^{-1}({\hat z}\times \nabla \psi)\cdot \nabla. 
\end{equation}   

The perpendicular electron velocity is more complicated. It has two parts, the E$\times$B drift and the polarization drift,
\begin{equation}
{\bf v}_\perp={\bf v}_E -\frac{m_\mathrm{e} }{e B^2}\,{\bf B}\times \frac{d_E}{dt}{\bf v}_E.
\label{eq:vperp}
\end{equation} 
In this expression, ${\bf v}_E=({\bf E}\times{\bf B})/B^2$ is the E$\times$B drift velocity, the convective derivative is $d_E/dt\equiv \partial/\partial t+{\bf v}_E\cdot\nabla$, and we note that $B^2\approx B_0^2+2B_0\delta B_z$. The second term in the right hand side of Equation (\ref{eq:vperp}) is smaller than the first by $\omega/\Omega_\mathrm{e}$, however, it needs to be kept since in the continuity equation the leading contribution from the first term cancels out. The electric field has both potential and non-potential components, ${\bf E}=-\nabla \phi+{\bf E}_\mathrm{np}$. The non-potential component is small compared to the potential one as it includes the (small) time derivative of the (small) magnetic fluctuations,
\begin{equation}
\nabla \times {\bf E}_\mathrm{np}=-\frac{\partial {\bf \delta B}}{\partial t},
\end{equation}
however, it also needs to be kept for the same reason. The fluctuations of the perpendicular magnetic field component are given by the $z$ component of the magnetic vector potential, $\delta\mathbf{B}_\perp={\hat z}\times\nabla\psi$. 

Substituting these expressions into Equation (\ref{eq:econteq}), we obtain
\begin{multline}
\frac{\partial}{\partial t}\left(\frac{\delta n}{n_0}-\frac{\delta B_z}{B_0}+\frac{m_\mathrm{e}}{eB_0^2}\nabla^2\phi\right)\\
+\frac{1}{B_0}\left({\hat z}\times \nabla \phi \right)\cdot \nabla \left(\frac{\delta n}{n_0}-\frac{\delta B_z}{B_0}+\frac{m_\mathrm{e}}{eB_0^2}\nabla^2\phi\right)\\
-\frac{1}{\mu_0 n_0 e}\nabla_\|\nabla_\perp^2\psi=0.
\label{eq:econteqlong}
\end{multline}

We now turn to the electron momentum equation. The parallel component of the equation is
\begin{equation}
\frac{\partial v_\|}{\partial t}+ ({\bf v}_E \cdot \nabla) v_\|=-\frac{e}{m_\mathrm{e}}E_\| -\frac{1}{m_\mathrm{e} n}\nabla_\| p_\mathrm{e}.
\end{equation}
Here, we have made use of $k_\perp\gg k_\|$ and $v_\|\sim v_\perp$. The latter condition can be checked a posteriori from the expressions for the parallel and perpendicular velocity components. The electric field is expressed through the scalar and vector potentials, 
\begin{equation}
E_\|=-\nabla_\|\phi-\partial A_\|/\partial t.
\end{equation}
As can be checked, the electron pressure fluctuations can be neglected compared to the electric potential fluctuations, $\phi$, for $T_\mathrm{e}\ll T_\mathrm{i}$. We then get the equation for the $z$ component of the magnetic potential $\psi$, 
\begin{equation}
\frac{\partial}{\partial t}\left(\psi-d_\mathrm{e}^2\nabla_\perp^2\psi \right)-d_\mathrm{e}^2\left({\bf v}_E\cdot \nabla\right)\nabla_\perp^2\psi=\nabla_\|\phi.
\label{eq:phi}
\end{equation}

To close Equations (\ref{eq:econteqlong}) and (\ref{eq:phi}), we need to find two additional relations among the fluctuating fields. One of these can be found from the perpendicular component of the electron momentum equation. Neglecting the small electron pressure, the force balance in the perpendicular direction is
\begin{eqnarray}
{\bf E}_\perp + {\bf v}_\perp \times {\bf B}=0.
\end{eqnarray}
Writing the perpendicular velocity as ${\bf v}_\perp=-{\bf J}_\perp /(ne)=-(\nabla \times {\bf B})_\perp/(\mu_0 n e)$, the electric field is then given by
\begin{eqnarray}
{\bf E}_\perp \approx \left[\left({\bf B}\cdot \nabla\right){\delta \bf B}_\perp-B_0\nabla_\perp \delta B_z \right]/(\mu_0 ne). 
\label{eq:eperp}
\end{eqnarray}
Since, as we will see later, $\delta B_z\sim \delta B_\perp$, the first term in the right hand side of Equation (\ref{eq:eperp}) is small. Using $k_\perp\gg k_\|$, the relation between the electric potential and the $z$ component of the fluctuating magnetic field is obtained,
\begin{eqnarray}
\phi = \frac{B_0}{\mu_0 n e}\delta B_z.
\label{eq:phi2}
\end{eqnarray} 
For the second relation, we use the fact that kinetic Alfv\'en fluctuations exist for $\omega \ll kv_{\mathrm{th},\mathrm{i}}$. 
In this case, the ion (and, by quasineutrality, electron) density fluctuations adjust to the electric potential according to the Boltzmann formula,
\begin{eqnarray}
\phi=-\frac{T_\mathrm{i}}{e}\frac{\delta n}{n_0}.
\label{eq:boltzmann}
\end{eqnarray}
It can be seen that Equations (\ref{eq:phi2}) and (\ref{eq:boltzmann}) agree with the force balance in the single-fluid momentum equation, ${\delta B_z}/{B_0}=-({\beta_\mathrm{i}}/{2})({\delta n}/{n_0})$. We can now eliminate the $\delta B_z$ and $\phi$ fluctuations from Equations (\ref{eq:econteqlong}) and (\ref{eq:phi}) in favor of the density fluctuations. In the following, we use the dimensionless variables, $t'=t |\Omega_\mathrm{e}|$, ${\bf x}'={\bf x}/d_\mathrm{e}$, $\psi'=\psi/(d_\mathrm{e} B_0)$, $\delta n'=(\beta_\mathrm{i}/2)(\delta n/n_0)$, and omit the prime signs. Keeping the leading-order terms, we obtain, after somewhat lengthy manipulations, the nonlinear system of equations for inertial kinetic Alfv\'en turbulence:
\begin{equation}
\frac{\partial}{\partial t}\left(1-\nabla_\perp^2\right)\psi+\left[({\hat z}\times\nabla\delta n)\cdot\nabla\right]\nabla_\perp^2\psi=-\nabla_\|\delta n, 
\label{eq:final1}
\end{equation}
\begin{equation}
\frac{\partial}{\partial t}\left(1+\frac{2}{\beta_\mathrm{i}}-\nabla_\perp^2\right)\delta n+\left[({\hat z}\times\nabla\delta n)\cdot\nabla\right]\nabla_\perp^2\delta n=\nabla_\|\nabla_\perp^2\psi.
\label{eq:final2}
\end{equation}

The nonlinearities in these equations appear in the second terms on the left-hand sides, and in the parallel gradients on the right-hand sides, 
\begin{equation}
\nabla_\|=\partial/\partial z+({\hat z}\times \nabla \psi)\cdot \nabla .
\end{equation}
Without the nonlinear terms, these equations reproduce the inertial kinetic Alfv\'en wave dispersion relation (Equation (\ref{eq:ikaw2})), and the relation between $\delta n$ and $\psi$ for these modes,
\begin{equation}
\delta n=\sqrt{\frac{1+k_\perp^2}{1+2/\beta_\mathrm{i}+k_\perp^2}}k_\perp\psi.
\end{equation}
The last equation agrees with Equation (\ref{eq:ikawcompressibility}) since $k_\perp\psi$ represents the perpendicular magnetic fluctuations, and $\delta n=-\delta B_z$ in the above normalization.

\subsection{Energy Spectrum and Anisotropy}
\label{sec:spectrumanisotropy}

From Equations (\ref{eq:final1}) and (\ref{eq:final2}), we can derive the spectrum of inertial kinetic Alfv\'en turbulence. In the absence of energy supply and dissipation, the equations conserve the energy
\begin{equation}
E=\int\left[\delta n\left(1+\frac{2}{\beta_\mathrm{i}}-\nabla_\perp^2 \right)\delta n-\nabla_\perp^2\psi\left(1-\nabla_\perp^2\right)\psi\right]d^3\mathbf{x}.
\label{eq:energy}
\end{equation}
In a turbulent state, both terms of $E$ are of the same order. For scales $k_\perp^2\gg1+2/\beta_\mathrm{i}$, this means that $\delta n_\lambda\sim\psi_\lambda/\lambda$, where $\delta n_\lambda$ and $\psi_\lambda$ denote the typical fluctuations of the fields at the scale $\lambda$ across the background magnetic field. In the same limit, the nonlinearity is dominated by the terms in the left-hand sides of Equations (\ref{eq:final1}) and (\ref{eq:final2}), and the nonlinear time can be estimated as $\tau\sim\lambda^2/\delta n_\lambda$.   Assuming a constant energy flux through scales, $\varepsilon\sim(\delta n_\lambda^2/\lambda^2)/\tau\sim\delta n_\lambda^3/\lambda^4$, leads to the scaling of the density and magnetic fluctuations $\delta n_\lambda\sim\psi_\lambda/\lambda\sim\varepsilon^{1/3}\lambda^{4/3}$. Applying the Fourier transform in the plane perpendicular to the background magnetic field, the spectrum of density and perpendicular fluctuations is obtained, $E_{n,B}(k_\perp)\propto k_\perp^{-11/3}$. In the data interval discussed in Section \ref{sec:observations}, the scaling range between $kd_\mathrm{e}\sim 1$ and $k\rho_\mathrm{i}\sim 1$ is not large (a factor of 3.4), so a well-developed power law spectrum is not present, but the measured spectral index can still be compared to the prediction in the limited range. Figure \ref{fig:compressibility} shows the spectral index to be $-3.6$ for the $\delta B_\perp$ fluctuations between $kd_\mathrm{e}=1$ and $k\rho_\mathrm{e}=1$, which is not far from the predicted value of $-11/3$.   

The anisotropy implied by the critical balance condition can also be determined. Balancing the linear and nonlinear terms in Equations (\ref{eq:final1}) and (\ref{eq:final2}), $\psi_\lambda/(l_\|\lambda^2)\sim \delta n_\lambda^2/\lambda^4$, we obtain the relation between the parallel and perpendicular scales $l_\|\sim \lambda^{5/3}$. In Fourier space this means that the turbulent energy is concentrated in the region $k_\|\lesssim k_\perp^{5/3}$, which becomes progressively broader in $k_\|$ and less anisotropic as $k_\perp$ increases. This suggests, therefore, that in contrast to standard Alfv\'en and kinetic Alfv\'en turbulence, the energy cascade in inertial kinetic Alfv\'en turbulence supplies energy more efficiently to $k_\|$ rather than $k_\perp$ modes. This anisotropy also implies that the Taylor condition becomes better satisfied as $k_\perp$ increases (see Section \ref{sec:taylor}). The current data interval does not allow the scale-dependence of the anisotropy to be tested, but this could be done in future studies with larger data sets, and also tested in numerical simulations.

\subsection{Inertial Whistler Turbulence}

The condition derived in Section \ref{sec:spectrumanisotropy} that critically balanced inertial kinetic Alfv\'en turbulence becomes more isotropic towards smaller scales leads to the interesting possibility that the cascade may transition to inertial whistler turbulence if the anisotropy reduces sufficiently. This is because there is a maximum value of $k_\|/k_\perp$ at which inertial kinetic Alfv\'en waves can exist (see Appendix \ref{sec:ikawderivation}); the broadening of the spectrum in the $k_\|$ direction would lead to the turbulence reaching $\omega>kv_{\mathrm{th},\mathrm{i}}$, which is the whistler frequency range.

For completeness, we give here the dynamical equations for inertial whistler turbulence. The basic equations (\ref{eq:econteqlong}) and (\ref{eq:phi}) hold for whistler turbulence as well, however, the additional condition (\ref{eq:boltzmann}) needs to be modified. Due to the high frequency of the whistlers, the density fluctuations do not follow the electric potential and remain small, 
\begin{equation}
\frac{\delta n}{n_0}\ll -\frac{e\phi}{T_\mathrm{i}}.
\end{equation}
We can thus neglect the density fluctuations in Equations (\ref{eq:econteqlong}) and (\ref{eq:phi}), and use Equation (\ref{eq:phi2}) to remove the electric potential fluctuations. Using again the normalized variables, and also $\delta B_z'=\delta B_z/B_0$, with the primes omitted, we obtain the nonlinear equations for inertial whistler turbulence:
\begin{equation}
\frac{\partial}{\partial t}\left(1-\nabla_\perp^2\right)\psi-\left[({\hat z}\times\nabla\delta B_z)\cdot\nabla\right]\nabla_\perp^2\psi=\nabla_\|\delta B_z, 
\label{eq:final3}
\end{equation}
\begin{equation}
\frac{\partial}{\partial t}\left(1-\nabla_\perp^2\right)\delta B_z-\left[({\hat z}\times\nabla\delta B_z)\cdot\nabla\right]\nabla_\perp^2\delta B_z=-\nabla_\|\nabla_\perp^2\psi.
\label{eq:final4}
\end{equation}
Linearization of these equations leads to the inertial whistler wave dispersion relation, which (in unnormalized variables) is given by
\begin{equation}
\omega^2=\frac{k_\|^2v_\mathrm{A}^2k_\perp^2\rho_\mathrm{i}^2}{\beta_\mathrm{i}(1+k_\perp^2d_\mathrm{e}^2)^2},
\end{equation}
and is also derived in Appendix \ref{sec:iwderivation}.

Due to the structural similarity of the inertial kinetic Alfv\'en and inertial whistler equations, (\ref{eq:final1}, \ref{eq:final2}) and (\ref{eq:final3}, \ref{eq:final4}), the spectrum of magnetic fluctuations in inertial whistler turbulence is the same as that derived in Section \ref{sec:spectrumanisotropy}, with the difference that the transition to the inertial regime occurs at $k_\perp^2 > 1$ rather than $k_\perp^2 > 1+2/\beta_\mathrm{i}$. This spectrum for inertial whistler turbulence has also been discussed previously \citep{biskamp96,biskamp99a,meyrand10,andres14}. The structural similarity between the equations means that the anisotropy implied by the critical balance condition (discussed in Section \ref{sec:spectrumanisotropy}) is also the same, meaning that the increasing isotropization would continue if the inertial kinetic Alfv\'en cascade transitions to inertial whistler turbulence.

The main physical difference between inertial kinetic Alfv\'en and inertial whistler turbulence is the ion dynamics, as discussed above, which leads to negligibly small density fluctuations in inertial whistler turbulence (see Appendix \ref{sec:iwderivation}). The magnetic compressibility, however, is the same, with $\delta B_\perp^2/\delta B_\|^2=1$ for both types of turbulence at sub-electron-inertial scales. The measurements in Figure \ref{fig:compressibility}, therefore, allow for the possibility of a transition to inertial whistler turbulence between $kd_\mathrm{e}\sim 1$ and $k\rho_\mathrm{e}\sim 1$. Further measurements will be required determine whether, and under what conditions, such a transition occurs.

\section{Discussion}

We have presented measurements of kinetic scale turbulence in the Earth's magnetosheath, which have the conditions $T_\mathrm{i}\gg T_\mathrm{e}$ and $\beta_\mathrm{i}\sim 1$. In the first decade of the kinetic range, the turbulence is similar to that in the upstream solar wind: it is predominantly low-frequency ($\omega\ll k_\perp v_{\mathrm{th},\mathrm{i}}$), anisotropic ($k_\perp\gg k_\|$), and kinetic Alfv\'en in nature, with spectra that match theoretical predictions and numerical simulations. In the second decade, however, a regime of inertial kinetic Alfv\'en turbulence has been identified by the increase in magnetic compressibility following that of the inertial kinetic Alfv\'en wave (Equation (\ref{eq:ikawcompressibility})). A set of nonlinear equations (Equations (\ref{eq:final1})-(\ref{eq:final2})) has been derived, which can be used to obtain the spectrum of magnetic fluctuations, $E_{B}(k_\perp)\propto k_\perp^{-11/3}$, between the electron inertial scale and electron gyroscale, which is consistent with the observed spectral steepening. Interestingly, this turbulence is expected to exhibit a qualitatively different scale-dependent anisotropy to standard Alfv\'en and kinetic Alfv\'en turbulence, becoming less anisotropic towards smaller scales. This increasing isotropisation may lead to a transition to inertial whistler turbulence (described by Equations (\ref{eq:final3}) and (\ref{eq:final4})) if the frequency reaches $\omega>k_\perp v_{\mathrm{th},\mathrm{i}}$. We plan to investigate these aspects with further observations and numerical simulations.

As well as in the Earth's magnetosheath, inertial kinetic Alfv\'en turbulence may also be present in several other astrophysical environments, with comparable plasma parameters. For example, the fast solar wind model of \citet{chandran11} predicts $\beta_\mathrm{i}\sim 0.1$ and $\beta_\mathrm{e}\sim 0.02$ at 10 Solar radii from the Sun, a regime in which inertial kinetic Alfv\'en turbulence would be expected to constitute a significant fraction of the kinetic range. This region of the solar corona will soon be measured \emph{in situ} by the \emph{Solar Probe Plus} spacecraft \citep{fox16}, allowing this to be tested directly. Similarly, in hot accretion flows, where turbulent heating is thought to be important, the ions are likely to be hotter than the electrons \citep{quataert98}, allowing the possibility of inertial kinetic Alfv\'en turbulence at small scales. Finally, collisionless shocks, such as the one that generates the Earth's magnetosheath, are common throughout the universe, leading to turbulent regions of space with large $T_\mathrm{i}/T_\mathrm{e}$ \citep{treumann09,ghavamian13}. Inertial kinetic Alfv\'en turbulence, therefore, may be quite widespread, and a possible route through which astrophysical plasmas are heated.

\acknowledgments
C.H.K.C. is supported by an STFC Ernest Rutherford Fellowship. S.B. is supported by the Space Science Institute, NSF grant AGS-1261659, and by the Vilas Associates Award from UW Madison. We acknowledge the \emph{MMS} team for producing the data, which was obtained from the \emph{MMS} Science Data Center ({https://lasp.colorado.edu/mms/sdc/}).

\appendix

\section{Derivation of the Inertial Kinetic Alfv\'en Wave}
\label{sec:ikawderivation}

In a $\beta\sim 1$ plasma, when the wave propagation is oblique, $k_\perp \gg k_\|$, the Alfv\'en wave transforms into the kinetic Alfv\'en wave for scales $k_\perp\rho_\mathrm{i} >1$. The situation is different, however, when either $\beta_\mathrm{i}$ or $\beta_\mathrm{e}$ are small. In the regime of extremely small electron plasma beta $\beta_\mathrm{e} \ll m_\mathrm{e}/m_\mathrm{i} $, and $\beta_\mathrm{i}\ll 1$, the Alfv\'en wave transforms into the inertial Alfv\'en wave \cite[e.g.,][]{lysak96}. In this case, the electron thermal velocity is much smaller than the Alfv\'en velocity, and the electrons do not adjust instantaneously to the electric field acting along the magnetic field lines. The regime here, however, is different, $m_\mathrm{e}/m_\mathrm{i} \ll \beta_\mathrm{e} \ll 1$, and the standard inertial Alfv\'en theory does not apply.

The dispersion relation for the standard kinetic Alfv\'en wave \cite[e.g.,][]{howes06}, derived under the assumption $\omega \ll k_z v_{\mathrm{th},\mathrm{e}}$, is
\begin{equation}
\omega^2=\frac{k_z^2v_\mathrm{A}^2k_\perp^2\rho_\mathrm{i}^2(1+T_\mathrm{e}/T_\mathrm{i})}{2+\beta_\mathrm{i}(1+T_\mathrm{e}/T_\mathrm{i})}\approx\frac{k_z^2v_\mathrm{A}^2k_\perp^2\rho_\mathrm{i}^2}{2+\beta_\mathrm{i}},
\label{eq:kaw2}
\end{equation}
where $T_\mathrm{e}\ll T_\mathrm{i}$ has been assumed in the last expression. Note that in the linear theory, the global and local mean magnetic field directions are the same, and here we use $z$ for this direction. It can be seen, however, that the frequency of the kinetic Alfv\'en wave becomes larger than $k_zv_{\mathrm{th},\mathrm{e}}$ for
\begin{equation}
k_\perp^2\rho_\mathrm{e}^2>\beta_\mathrm{i}\left(2+\beta_\mathrm{i}\right)\left(T_\mathrm{e}/T_\mathrm{i}\right)^2.
\end{equation}
For $T_\mathrm{e}\ll T_\mathrm{i}$ and $\beta_\mathrm{i}\sim 1$, this happens at a scale much larger than the electron gyroscale.

In the frequency range $k_zv_{\mathrm{th},\mathrm{e}}\ll \omega \ll k_\perp v_{\mathrm{th},\mathrm{i}}$, the kinetic Alfv\'en wave transforms into the inertial kinetic Alfv\'en wave. To derive this mode, we consider a collisionless plasma with $\beta_\mathrm{i}\lesssim 1$, at sub-ion scales $k_\perp\rho_\mathrm{i}\gg 1$. The wave modes can be found from the equation $D_{ij}E_{j}=0$, where $D_{ij}=k^2\delta_{ij}-k_ik_j-(\omega^2/c^2)\epsilon_{ij}$, $\epsilon_{ij}$ is the plasma dielectric tensor, and $E_j$ is the electric field. Under the additional assumption $T_\mathrm{e}\ll T_\mathrm{i}$, the components of the tensor $D_{ij}$ take the form
\begin{eqnarray}
D_{xx}&\approx & k_z^2-\frac{\omega^2}{c^2}\frac{\omega_\mathrm{pe}^2}{\Omega_\mathrm{e}^2}\left(1+\frac{2}{\beta_\mathrm{i} k_\perp^2 d_\mathrm{i}^2} \right), \label{eq:dxx}\\
D_{xy}&= & -D_{yx}\approx -i\frac{\omega\omega_\mathrm{pe}^2}{\Omega_\mathrm{e} c^2},\\
D_{xz}&= & D_{zx}\approx -k_zk_\perp,\\
D_{yy}&\approx&k^2,\\
D_{zz}&\approx&k_\perp^2+\frac{\omega_\mathrm{pe}^2}{c^2},\\
D_{yz}&=&D_{zy}\approx 0,\label{eq:dyz}
\end{eqnarray}
where the wave vector has components $(k_\perp,0,k_z)$.

The dispersion relation for the inertial kinetic Alfv\'en wave obtained from this equation is
\begin{equation}
\omega^2=\frac{k_z^2 k_\perp^2d_\mathrm{e}^4 \Omega_\mathrm{e}^2}{\left(1+k_\perp^2d_\mathrm{e}^2\right)\left(1+{2}/{\beta_\mathrm{i}}+k_\perp^2d_\mathrm{e}^2 \right)}.
\label{eq:ikaw2}
\end{equation}
When the electron inertial corrections are small, $k_\perp^2 d_\mathrm{e}^2 \ll 1$, this is similar to the dispersion relation of the standard kinetic Alfv\'en wave (Equation (\ref{eq:kaw2})), although in a different phase space region, $\omega\gg k_z v_{\mathrm{th},\mathrm{e}}$. In the opposite case, $k_\perp^2 d_\mathrm{e}^2 \gg 1+2/\beta_\mathrm{i}$, the inertial kinetic Alfv\'en wave frequency becomes
\begin{equation}
\omega^2 = \Omega_\mathrm{e}^2{k_z^2}/{k_\perp^2}.
\end{equation}
A similar derivation for the Alfv\'en modes at low $\beta$ can also be made starting from the gyrokinetic description \citep[e.g.,][]{howes06,zocco11}.

The inertial kinetic Alfv\'en wave exists for $\omega^2 \ll k_\perp^2v^2_{\mathrm{th},\mathrm{i}}$, which, together with the dispersion relation (Equation (\ref{eq:ikaw2})), allows the required obliquity to be determined. At $k_\perp^2 d^2_\mathrm{e}\sim 1$ the wave requires $k^2_z/k^2_\perp \ll {m_\mathrm{e}/m_\mathrm{i}}$, and at $k_\perp^2 \rho^2_\mathrm{e}\sim 1$ the condition is $k^2_z/k^2_\perp \ll ({m_\mathrm{e}/m_\mathrm{i}})(T_\mathrm{i}/T_\mathrm{e})$.

The density fluctuations associated with the inertial kinetic Alfv\'en wave are 
\begin{equation}
\left(\frac{\delta n}{n_0} \right)^2=\frac{4}{\beta_\mathrm{i}^2}\left(1+\frac{1+2/\beta_\mathrm{i}+k_\perp^2d_\mathrm{e}^2}{1+k_\perp^2d_\mathrm{e}^2} \right)^{-1}\left( \frac{\delta {\bf B}}{B_0}\right)^2,
\label{eq:ikawdb}
\end{equation}
and they are anti-correlated with the fluctuations of the magnetic field strength,
\begin{equation}
\frac{\delta B_z}{B_0}=-\frac{\beta_\mathrm{i}}{2} \frac{\delta n}{n_0}.
\label{eq:ikawdbz}
\end{equation}
From Equations (\ref{eq:ikawdb}) and (\ref{eq:ikawdbz}), the expression for the magnetic compressibility is obtained,
\begin{equation}
\frac{\delta B_z^2}{\delta B_\perp^2}=\frac{1+k_\perp^2d_\mathrm{e}^2}{1+2/\beta_\mathrm{i}+k_\perp^2d_\mathrm{e}^2}.
\end{equation}

\section{Derivation of the Inertial Whistler Wave}
\label{sec:iwderivation}

The whistler wave, unlike the kinetic Alfv\'en wave, exists for frequencies $\omega>k_\perp v_{\mathrm{th},\mathrm{i}}$ and transforms into the inertial whistler wave at $k_\perp d_\mathrm{e}\sim 1$. It can be derived from the same equations~(\ref{eq:dxx}-\ref{eq:dyz}), if $D_{xx}$ is replaced by
\begin{equation}
D_{xx}=k_z^2 -\frac{\omega^2 \omega_\mathrm{pe}^2}{c^2 \Omega_\mathrm{e}^2}+ \frac{\omega_\mathrm{pi}^2}{c^2}.
\end{equation}
The dispersion relation for the inertial whistler wave is then obtained,
\begin{equation}
\omega^2=k_z^2k^2d_\mathrm{e}^4\Omega_\mathrm{e}^2\frac{\left[1+\frac{1+k_\perp^2d_\mathrm{e}^2}{k_z^2d_\mathrm{i}^2} \right]}{\left(1+k^2d_\mathrm{e}^2\right)\left(1+k_\perp^2d_\mathrm{e}^2 \right)}.
\label{eq:omegaw}
\end{equation}
Without the electron inertial corrections, $kd_\mathrm{e} \ll 1$, the wave exists only for $k_zd_\mathrm{i}\gg 1$, where we recover the well-known whistler dispersion relation. In the opposite limit, $k_\perp d_\mathrm{e} \gg 1$, the whistlers exist for $k_zd_\mathrm{i}\gg k_\perp^2 d_\mathrm{e}^2$, so the second term in the square bracket in Equation (\ref{eq:omegaw}) can always be neglected. {Equation (\ref{eq:omegaw}) then matches the previously studied whistler wave in the inertial regime \cite[e.g.,][]{biskamp96,biskamp99a}.}

The density fluctuations associated with the inertial whistler wave are given by
\begin{equation}
\left(\frac{\delta n}{n_0} \right)^2=\frac{\left(1+k_\perp^2d_\mathrm{e}^2\right)^4}{\left(1+k_z^2d_\mathrm{i}^2+k_\perp^2d_\mathrm{e}^2\right)^2\left[1+\frac{k_z^2d_\mathrm{i}^2}{1+k_z^2d_\mathrm{i}^2+k_\perp^2d_\mathrm{e}^2}\right]}\left(\frac{\delta {\bf B}}{B_0}\right)^2,
\end{equation}
and are positively correlated with the fluctuations of the magnetic field strength,
\begin{equation}
\frac{\delta n}{n_0}=\frac{(1+ k_\perp ^2d_\mathrm{e}^2)^2}{1+k_\perp^2 d_\mathrm{e}^2+k_z^2 d_\mathrm{i}^2}\frac{\delta B_z}{B_0}.
\end{equation}
For $kd_\mathrm{e}\ll 1$, the wave satisfies $k_zd_\mathrm{i}\gg 1$ and we recover the well-known relation, $(\delta n/n_0)^2/(\delta B_z/B_0)^2=1/(k_z^4d_\mathrm{i}^4)$. In the inertial regime $k_\perp d_\mathrm{e}\gg 1$ it satisfies $k_zd_\mathrm{i}\gg k_\perp^2d_\mathrm{e}^2$, and we obtain $(\delta n/n_0)^2/(\delta B_z/B_0)^2=(k_\perp^4 d_\mathrm{e}^4)^2/(k_z^2d_\mathrm{i}^2)^2$. In both cases, the density fluctuations are much smaller than the magnetic fluctuations. Also, in both cases the magnetic compressibility is given by
\begin{equation}
\frac{\delta B_z^2}{\delta B_\perp^2}=1.
\end{equation}

\bibliography{bibliography}

\end{document}